\documentclass[seceq]{ptptex}

\usepackage{graphicx}
\usepackage{wrapft}
\usepackage{psfrag}
\newcommand{\eq}[1]{Eq.~(\ref{#1})}
\def\ee{\mathrm{e}}

\def\one{{\rm 1\kern -.9mm l}}

\def\ii{{\mbox{i}}}

\newcommand{\uno}{\mbox{1\!\negmedspace1}}
\newcommand{\lvev}{\Big\langle\hskip -5pt\Big\langle}
\newcommand{\rvev}{\Big\rangle\hskip -5pt\Big\rangle}

\title{
$\mathcal{N}=2$ Instanton Calculus In Closed String Background
}

\author{
Marco \textsc{Bill\'o $^1$\footnote{E-mail: {\tt billo@to.infn.it}}},
Marialuisa \textsc{Frau $^1$\footnote{E-mail: {\tt frau@to.infn.it}}},
Alberto \textsc{Lerda $^2$\footnote{E-mail: {\tt lerda@to.infn.it}}}
}

\inst{$^1$ Dipartimento di Fisica Teorica, Universit\`a di Torino\\
and I.N.F.N. - sezione di Torino,
via P. Giuria 1, I-10125 Torino, Italy\\
$^2$ Dipartimento di Scienze e Tecnologie Avanzate\\
Universit\`a del Piemonte Orientale, I-15100 Alessandria, Italy\\
and I.N.F.N.  Gruppo Collegato di Alessandria - sezione di Torino
}

\abst{
In this contribution we describe how to obtain
instanton effects in four dimensional
gauge theories by computing string
scattering amplitudes in D3/D(--1) brane systems. In particular
we study a system of fractional D3/D(--1) branes in a $\mathbb{Z}_2$ orbifold and
in a Ramond-Ramond closed string background, and show that it describes the gauge instantons of
$\mathcal{N}=2$ super Yang-Mills theory and their interactions with the graviphoton of
$\mathcal{N}=2$. Using string theory methods we compute the prepotential of
the effective gauge theory exploiting
the localization methods of the instanton calculus, showing that this leads to the
same information given by the topological string.
}

\begin{document}

\maketitle

\section{Introduction}

The relationship between string theory and perturbative field theories
has been thoroughly investigated for many years.
The study of non-perturbative effects in string theory and their
comparison with field theory is instead much more recent.
In particular, only after the introduction of D branes it has been
possible to significantly improve our knowledge and address some non-perturbative
issues using string theory.
For instance, it has been shown~
\cite{Witten:1995im,douglas,Green:2000ke}
that the instanton sectors of
${\mathcal N}=4$ supersymmetric Yang-Mills (SYM) gauge theory
can be described in string theory by systems of
D3 and D(--1) branes (or D-instantons) in flat space. In fact, the excitations of the
open strings stretching between two D(--1) branes
or between a D3 brane and a D-instanton, are in one-to-one
correspondence with the moduli of the SYM instantons in the so-called
ADHM construction (for comprehensive reviews on the subject see, for instance,
Ref. ~\cite{Dorey:2002ik}).

The above remarks have been further substantiated~ \cite{Billo:2002hm} by showing
that the tree-level string scattering amplitudes on disks
with mixed boundary conditions for a D3/D(--1) system lead, in the
infinite tension limit $\alpha^\prime\to 0$, to the effective action on
the instanton moduli space of the SYM theory.
Furthermore, it has been proved that the same disk diagrams
also yield the classical field profile of the instanton solution, and that
these mixed disks effectively act as sources for the various components
of the gauge supermultiplet. This approach can be easily adapted to
describe SYM theories with $\mathcal{N}=2$ and $\mathcal{N}=1$ supersymmetry and their
instantons: instead of considering D3/D(--1) systems in flat space, one has simply to
place the branes at suitable orbifold singularities.

This description of gauge theories by means of systems of D-branes
has proven to be very useful also to account for the
deformations induced by non-trivial gravitational
backgrounds. For instance, non-commutative gauge theories and
their instanton sectors can be efficiently described
by considering D3/D(--1) brane systems in a Kalb-Ramond background~ \cite{Billo:2005fg} .
Similarly, by placing the branes in a particular R-R background, non-anti-commutative
gauge theories can be described and the corresponding non-anti-commutative instanton solutions
can be determined~ \cite{Billo:2004zq} .
More recently, the string description of instantons has lead to
new developments. In fact it has been shown in several different
contexts that these stringy instantons may dynamically generate new
types of terms in the low-energy effective action of the SYM
theory with very interesting phenomenological implications
~\cite{recent} .

In this contribution, which is mainly based on Ref.~ \cite{Billo:2006jm} ,
we will concentrate on ${\mathcal N}=2$ SYM theory in four dimensions
and study the non-perturbative low energy effective action induced by
instantons when a R-R graviphoton background is switched on.
We show how to determine the non-perturbative gravitational F-terms
in the ${\mathcal N}=2$ prepotential exploiting localization
methods~ \cite{Nekrasov:2002qd,Losev:2003py} .

The instanton sectors of ${\mathcal N}=2$ SYM theory with gauge group
$\mathrm{SU}(N)$ can be described by a system of $N$ fractional D3 branes
and $k$ fractional D-instantons in the orbifold
$\mathbb{R}^{4}\times \mathbb{C}\times\mathbb{C}^2/\mathbb{Z}_2$
where the orbifold group $\mathbb{Z}_2$ acts as the inversion of the last two complex
coordinates.
The gauge theory degrees of freedom are described by the massless excitations of the
open strings stretching between two D3 branes, and can be assembled into a ${\cal N}=2$ chiral
superfield in the adjoint of $\mathrm{SU}(N)$
\begin{equation}
\label{Pm1}
\Phi(x,\theta) = {\phi(x)} + \theta{\Lambda(x)} +\frac
12\,\theta\sigma^{\mu\nu} \theta
\,{F_{\mu\nu}^+(x)} + \cdots
\end{equation}
String amplitudes for the string vertices corresponding to the
SYM fields of \eq{Pm1} on {disks} attached to the {D3 branes}
give rise, in the limit {$\alpha'\to 0$} with {gauge quantities fixed},
to the tree level (microscopic) {$\mathcal{N}=2$} action for
$\mathrm{SU}(N)$ SYM.

We are interested in studying the low energy effective action
on the {Coulomb branch} parameterized by the {v.e.v.'s}
${\langle \Phi_{uv} \rangle} = {a_u}\,\delta_{uv}$
of the adjoint chiral superfields breaking
$\mbox{SU}(N) \to \mbox{U}(1)^{N-1}$.
{F}rom now on we will focus for simplicity on the $\mbox{SU}(2)$ theory broken to
$\mathrm{U}(1)$, and thus we will deal with a single chiral
superfield $\Phi$ and a single v.e.v. $a$.
Up to two-derivatives, $\mathcal{N}=2$ supersymmetry constrains
the effective action for {$\Phi$}
to be of the form
\begin{equation}
\label{Seff}
S_{\mbox{\tiny eff}}[{\Phi}] =
\int d^4x \,d^4\theta\, {\mathrm{F}}({\Phi}) + \mathrm{c.c}~,
\end{equation}
where the prepotential $\mathrm{F}(\Phi)$ is a holomorphic homogeneous function of
degree two.

The main purpose of our discussion is to show
how the instanton corrections to the prepotential arise in our
string set-up, also in presence of a non-trivial supergravity
background.
In particular we will show that the low energy excitations of the D(--1)/D(--1)
and D3/D(--1) open strings in the $\mathbb{Z}_2$ orbifold exactly account
for the ADHM instanton moduli of the $\mathcal{N}=2$ theory
and that the integration measure on
the moduli space is recovered from disk amplitudes.
Moreover, we show that the $\mathcal{N}=2$
super-instanton solution for the
fields appearing in \eq{Pm1} is obtained by computing one-point
open string amplitudes on mixed disks.
The fact that the instanton center
and its super-partners decouple from the  D3/D(--1) and D(--1)/D(--1) interaction
show that these moduli actually play the role of the $\mathcal{N}=2$ superspace coordinates;
therefore the instanton induced prepotential $\mathrm{F}(\Phi)$ may be identified with
the ``centered partition function'' of D-instantons coupled to
$\Phi$~ \cite{Fucito:1996ua,Hollowood:2002ds} .

This identification opens the way to several generalizations. In particular, one
may extend the above procedure to include also non-perturbative gravitational
terms in the effective action by computing the instanton partition function
in a non-trivial $\mathcal{N}=2$ supergravity background. In our context, this amounts to
compute the D(--1)/D(--1) and D3/D(--1) open string interactions in presence of
a closed string background, and obtain the deformed integration measure on the
instanton moduli space from mixed disk diagrams involving both open and closed
string vertices. In particular we will consider a $\mathcal{N}=2$ graviphoton background
with self-dual field strength $\mathcal{F}^+$ and determine the instanton induced
prepotential $\mathrm{F}(\Phi, W)$, where $W$ is the Weyl
superfield whose first component is related to the graviphoton
field strength $\mathcal{F}^+$. In this way we may obtain from mixed disk amplitudes
the gravitational F-terms of the form $(R^+)^2\,({\cal
F}^+)^{2h-2}$, where $R^+$ is the self-dual Riemann curvature tensor, which have
been previously computed~ \cite{Antoniadis:1993ze}
from topological string amplitudes at genus
$h$. Therefore, the application of localization
methods to the instanton calculus shows an interesting
relation~ \cite{Nekrasov:2002qd,Losev:2003py}
with the topological string which is worth further investigation.

\section{The D3/D(--1) system}

As we mentioned above, the $k$ instanton sector of a four-dimensional
SYM theory with gauge group
$\mathrm{SU}(N)$ can be described by a bound state of
$N$ D$3$ and $k$ D(--1) branes~ \cite{Witten:1995im} of fractional type in the space
${\mathbb{R}^4} \times \mathbb{M}^6$.
The amount of supersymmetry possessed by the SYM theory depends on the
features of the internal six-dimensional space $\mathbb{M}^6$. If we want
to describe ${\cal N}=2$ gauge theories, we can choose $\mathbb{M}^6$ to be the orbifold
$\mathbb{C} \times{\mathbb{C}^2/\mathbb{Z}_2}$.

In the D3/D(--1) system the string coordinates $X^{\cal
M}$ and $\psi^{\cal M}$ (${\cal
M}=1,\ldots,10$) obey different boundary conditions on
the two types of branes. Specifically, on the D(--1) branes we have
Dirichlet boundary conditions in all directions, while on the D3
branes the longitudinal fields $X^\mu$ and $\psi^\mu$ ($\mu=1,2,3,4$)
satisfy Neumann boundary conditions, and the transverse fields
$X^a$ and $\psi^a$ ($a=5,\ldots,10$) obey Dirichlet boundary
conditions.

The presence of the space-filling D3 branes and the orbifold projection break the
10-dimensional Euclidean invariance group
${\mbox{SO}(10)}$ into
$\mbox{SO}(4)\!\times\! \mbox{SO}(2)\!\times\! \mbox{SO}(4)_{\rm int}$
and therefore the 10-dimensional (anti-chiral) spin fields
$S^{\dot{\mathcal{A}}}$ decompose as follows
\begin{equation}
\label{sdec}
S^{\dot{\mathcal{A}}}\! \to\!
(S_{\alpha} S^{-} S_A, S_{\alpha} S^{+} S^{\dot A},
S^{\dot\alpha} S^{-} S^{\dot A}, S^{\dot\alpha} S^{+} S_A)~.
\end{equation}
where $S_\alpha$ ($S^{\dot\alpha}$) are $\mathrm{SO}(4)$ Weyl spinors of
positive (negative) chirality, $S_A$ ($S^{\dot A}$) are
$\mathrm{SO}(4)_{\rm int}$ Weyl
spinors of positive (negative) chirality and
$S^\pm$ are $\mathrm{SO}(2)$ spin fields with weight $\pm \frac 12$.
The chiral spin fields $S_A$ are even under the orbifold
generator, while the anti-chiral ones $S^{\dot A}$ are odd.

Since we consider a single type of fractional D3-branes, the Chan-Paton
factors for the open string excitations are invariant under the orbifold.
The orbifold projection therefore reduces the massless spectrum to the
bosonic states which arise by acting on the vacuum with the even fields
$\Psi= (\psi_5+{\rm i}\psi_6)/\sqrt{2}$ and
$\Psi^{\dagger}= (\psi_5-{\rm i}\psi_6)/\sqrt{2}$
and to the fermionic states associated with the invariant spin fields
$S_{\alpha} S^{-} S_A$ or $S^{\dot\alpha} S^{+} S_A$.

For our present purposes it is enough to specify the spectrum of
excitations of the open strings with at least one end-point on the
D-instantons, which, as explained in Ref.~ \cite{Billo:2002hm} , describe the
ADHM instanton moduli. Let us first consider the strings that have
both ends on the D($-1$) branes and therefore describe the neutral moduli:
in the NS sector the physical
excitations are $a_\mu$, $\chi$ and $\chi^{\dagger}$, whose corresponding
vertex operators are
\begin{equation}
\label{vertA}
V_a(z)={a^\mu}\,\psi_{\mu}(z)
\,{\rm e}^{-\phi(z)}~,~~
V_\chi(z)={\chi^{\dagger}}\,\Psi(z)\,{\rm e}^{-\phi(z)}~,~~
V_{\chi^{\dagger}}(z)={\chi}\,\Psi^{\dagger}(z)\,{\rm e}^{-\phi(z)}
\end{equation}
where $\phi$ is the chiral boson of the superghost bosonization.
In the R sector we find eight fermionic moduli which are conventionally
denoted by $M^{\alpha A}$ and $\lambda_{\dot\alpha A}$, and
correspond to the following vertex operators
\begin{equation}
\label{vertM'} V_{M}(z)\,=\, M^{\alpha A}\,
S_{\alpha}(z)S^-(z) S_A(z)\,{\rm e}^{-\frac{1}{2}\phi(z)}~,
~~
V_{\lambda}(z)\,=\,
{{\lambda_{\dot\alpha A}}}\,S^{\dot\alpha}(z) S^+(z)S^A(z)
\,{\rm e}^{-\frac{1}{2}\phi(z)}
~.
\end{equation}
Let us now consider the open strings that start on a D3 and end on a
D($-1$) brane, or vice-versa and describe the charged moduli.
These strings are characterized by the fact that the four directions
along the D3 branes have mixed boundary conditions.
Thus, in the NS sector the four
fields $\psi^\mu$ have integer-mode expansions with
zero-modes that represent the $\mathrm{SO}(4)$ Clifford algebra and
the corresponding physical excitations are organized in two
bosonic Weyl spinors of $\mathrm{SO}(4)$. These are denoted by
$w_{\dot\alpha}$ and $\bar w_{\dot\alpha}$ respectively, and are
described by the following vertex operators
\begin{equation}
\label{vertexw} V_w(z) \,=\,{w}_{\dot\alpha}\, \Delta(z)\,
S^{\dot\alpha}(z) \,{\rm e}^{-\phi(z)}~,~~
V_{\bar w}(z) \,=\,{\bar w}_{\dot\alpha}\, \bar\Delta(z)\,
S^{\dot\alpha}(z)\, {\rm e}^{-\phi(z)}~.
\end{equation}
Here $\Delta(z)$ and $\bar\Delta(z)$ are the bosonic twist and
anti-twist fields with conformal dimension $1/4$, that change the
boundary conditions of the $X^\mu$ coordinates from Neumann to
Dirichlet and vice-versa by introducing a cut in the world-sheet.
In the R sector the fields
$\psi^\mu$ have, instead, half-integer mode expansions
so that there are fermionic zero-modes only in the common
transverse directions. Thus, the physical excitations of this
sector form two fermionic Weyl spinors of ${\rm SO}(4)_{\rm int}$
which are denoted by $\mu^A$ and $\bar \mu^A$
respectively, and correspond to the following vertex operators
\begin{equation}
\label{vertexmu} V_\mu(z) \,=\,{\mu}^{A}\,
\Delta(z)\,S^-(z)S_{A}(z)\, {\rm e}^{-\frac{1}{2}\phi(z)}~~,~~
V_{\bar\mu}(z) \,=\,{{\bar \mu}}^{A}\,
\bar\Delta(z)\,S^-(z)S_{A}(z)\, {\rm e}^{-\frac{1}{2}\phi(z)}~.
\end{equation}
A systematic analysis~ \cite{Billo:2002hm} shows that, in the limit
$\alpha'\to 0$, the scattering amplitudes involving the above vertex
operators give rise to the following ``action''
\begin{eqnarray}
{S_{\rm mod}} = &
{\mbox{tr}_k}&\Big\{ -2\,[\chi^{\dagger},a'_\mu][\chi,{a'}^\mu] +
\chi^{\dagger}{\bar w}_{\dot\alpha} w^{\dot\alpha}\chi
+ \chi{\bar w}_{\dot\alpha} w^{\dot\alpha} \chi^{\dagger}
\nonumber\\
&&\!\!\!\!\!\!\!\!\!\!-\ii\,
\frac{\sqrt 2}{4}\,M^{\alpha A}\epsilon_{AB}[\chi^{\dagger},M_{\alpha}^{B}]
+ \ii\,
\frac{\sqrt 2}{2}\,{\bar \mu}^A \epsilon_{AB} \mu^B\chi^{\dagger}
\\
&&\!\!\!\!\!\!\!\!\!\!-\ii D_c \big({w_{\dot\alpha}(\tau^c)^{\dot\alpha}_{\dot\beta}
\bar{w}_{\dot\beta}
 +\ii
\bar\eta_{\mu\nu}^c \big[{a'}^\mu,{a'}^\nu\big]}\big) 
+ \ii {\lambda}^{\dot\alpha}_{\,A}\big(
{\bar{\mu}^A{w}_{\dot\alpha}+
\bar{w}_{\dot\alpha}{\mu}^A  +
\big[a'_{\alpha\dot\alpha},{M'}^{\alpha A}\big]\big)}\!
\Big\}~.
\nonumber
\label{smod}
\end{eqnarray}
where ${\rm tr}_k$ means trace over ${\mathrm U}(k)$,
$\bar\eta^c$ ($c=1,2,3$) are the anti-self dual 't Hooft symbols,
and $\tau^c$ are the Pauli matrices. In (\ref{smod})
there appear also three auxiliary fields $D_c$ whose string representation
is provided by the following vertex operators~ \cite{Billo:2002hm}
\begin{equation}
\label{vertaux}
V_D(z) \,=\, \frac{1}{2}
D_c\,\bar\eta_{\mu\nu}^c\,\psi^\nu(z) \psi^\mu(z)~.
\end{equation}
As is well known~ \cite{Dorey:2002ik} , by simply taking ${\rm e}^{-S_{\rm mod}}$
one obtains the measure on the instanton moduli space,
while by varying $S_{\rm mod}$ with respect to $D_c$ and
${\lambda}^{\dot\alpha}_{~A}$ one finds the bosonic and fermionic
ADHM constraints.

Notice that $S_{\rm mod}$ does not depend on the instanton center $x_0^{\mu}$
nor on its super-partners $\theta^{\alpha A}$, which are, respectively, the $U(1)$ components
of $a^{\mu}$ and $M^{\alpha A}$, namely
\begin{equation}
\begin{aligned}
{a'}^\mu &= x_0^\mu\,\uno_{[k]\times[k]} + y^\mu_c\,T^c~~,
\\
{M}^{\alpha A}&=\theta^{\alpha A}\,\uno_{[k]\times[k]} + {\zeta}^{\alpha
A}_c\,T^c~.
\end{aligned}
\label{xtheta}
\end{equation}
The moduli $x_0^{\mu}$ and $\theta^{\alpha A}$ actually play the role of
the $\mathcal{N}=2$ superspace coordinates, while the remaining moduli,
collectively denoted by ${\widehat{\cal M}_{(k)}}$ are the so-called ``centered
moduli''. Using this decomposition, the $k$-instanton partition function can
then be written as
\begin{equation}
Z_{(k)}= \int d^4x_0\, d^4\theta \,{\widehat Z_{(k)}}
\label{z}
\end{equation}
where
\begin{equation}
{\widehat Z_{(k)}}= \int d{\widehat{\cal M}_{(k)}}\,
{\rm e}^{-\frac{8\pi^2k}{g^2}-S_{\rm mod}({\widehat{\cal M}_{(k)}})}
\label{zk}
\end{equation}
is the centered partition function.
It is worth pointing out that among the ``centered moduli''
$\widehat{\cal M}_{(k)}$ there is the singlet part of the anti-chiral fermions
$\lambda_{\dot\alpha A}$ which is associated to the supersymmetries that are preserved
both by the D3 and by the D(--1) branes \footnote{This is to be contrasted with the
$\theta^{\alpha A}$ defined in (\ref{xtheta}), which are associated to the
supersymmetries preserved by the
D3 but broken by the D(--1) branes.}. Thus, despite the suggestive notation of
Eq. (\ref{z}), one may naively think that the full D-instanton partition function
cannot yield an F-term in the effective action, {\it i.e.}
an integral on half superspace, due to the presence of the anti-chiral
$\lambda_{\dot\alpha A}$'s among
the integration variables. Actually, this is not true since the
$\lambda_{\dot\alpha A}$'s, including its singlet part, do couple
to other instanton moduli (see the last terms in Eq. (\ref{smod}))
and their integration correctly enforces the fermionic ADHM constraints on
the moduli space. Therefore, the instanton
partition function (\ref{z}) does indeed yield non-perturbative
F-terms. Things are very different instead for the exotic
instantons that have been recently considered in the literature~
\cite{recent} . In this case, due to the different structure of the
charged moduli, the $\lambda_{\dot\alpha A}$'s do not couple to
anything and in order to get a non-vanishing result, they have to
be removed from the spectrum, for example with an orientifold projection.

\section{The instanton solution from mixed disks}

The construction of the ADHM instanton moduli and of their
integration measure in terms of open strings given in the previous
section clearly shows that gauge theory instantons are described
by systems of D3/D($-1)$ branes.
There is however an even more convincing evidence in favor of
this identification, namely the fact that the mixed disks
of the D3/D($-1)$ brane system are the
source for the instanton background of the super Yang-Mills theory,
and that the classical instanton profile can be obtained from open
string amplitudes.

\begin{figure}[t]
\begin{center}
\psfrag{mu}{ $A_\mu^c$}
\psfrag{I}{$ $}
\psfrag{p}{\small $p$}
\psfrag{w}{\small $\bar w$}
\psfrag{wb}{\small $w$}
\includegraphics[width=0.31\textwidth]{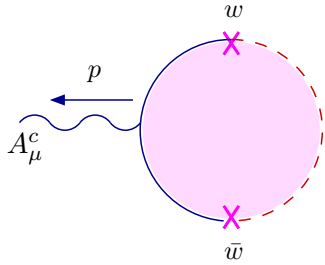}
\end{center}
\caption{The mixed disk that describes the emission of a gauge
vector field $A_\mu^c$ with momentum $p$ represented by the
outgoing wavy line.}\label{fig:md2}
\end{figure}

For simplicity we will discuss only the case of instanton number $k=1$
in a $\mathrm{SU}(2)$ gauge theory, but no substantial changes
occur in our analysis if one considers higher values of $k$ and
other gauge groups (see Ref.~ \cite{Billo:2002hm} for these
extensions). Let us then consider the emission of the
$\mathrm{SU}(2)$ gauge vector field $A_\mu^c$ from a mixed disk.
The simplest diagram which can contribute to this process contains
two boundary changing operators $V_{\bar w}$ and $V_{w}$ and no
other moduli, and is shown in Fig. \ref{fig:md2}.

\noindent
The amplitude associated to this diagram is
\begin{equation}
\label{dia1}
\left\langle {\cal V}_{A^c_\mu(-p)}\right\rangle_{\rm mixed~disk}
 \,\equiv\,
\lvev V_{\bar w}\,{\cal V}_{A^c_\mu(-p)}\,V_{w}
\rvev \,=\,  {\cal A}^c_\mu(p ;{\bar w, w})
\end{equation}
where ${\cal V}_{A^c_\mu(-p)}$ is the gluon vertex operator
with {outgoing} momentum $p$ and {without} polarization, namely
\begin{equation}
\label{vert2}
{\cal V}_{A^c_\mu(-p)}
= 2\ii\, T^c\left(
\partial X_\mu \,-\, \ii \,p\cdot \psi\, \psi_\mu\right)\,
\ee^{-\ii p\cdot X(z)}
\end{equation}
where $T^c$ is the adjoint $\mathrm{SU}(2)$ generator.
Note that that the amplitude (\ref{dia1}) carries the
structure and the quantum numbers of an emitted gauge vector field
of $\mathrm{SU}(2)$. The evaluation of the
amplitude (\ref{dia1}) is quite straightforward and the result
is~ \cite{Billo:2002hm}
\begin{equation}
\label{ampl}
{\cal A}^c_\mu(p ;{\bar w, w}) = {\rm i}\rho^2\, p^\nu\,\bar\eta^c_{\nu\mu}
\,{\rm e}^{-{\rm i} p\cdot x_0}
\end{equation}
where we have defined $\bar w^{\dot\alpha} w_{\dot\alpha}=2\rho^2$.
By taking the Fourier transform of (\ref{ampl}), after
inserting a free propagator, we obtain
\begin{equation}
{\cal A}_\mu^c(x)
\,=\,\int\frac{d^4p}{(2\pi)^2}\,\frac{1}{p^2}\,{\cal A}^c_\mu(p ;{\bar w, w})
\,{\rm e}^{{\rm i} p\cdot x} \,=\,
\frac{2\rho^2\,\bar\eta^c_{\mu\nu}\,(x-x_0)^\nu}{(x-x_0)^4}~~.
\label{solution}
\end{equation}
Eq. (\ref{solution}) is the leading term in the
large distance expansion ({\it i.e.} $|x-x_0|\gg\rho$) of the
$\mathrm{SU}(2)$ instanton
with center $x_0$ and size $\rho$ in the
\emph{singular gauge}, namely
\begin{equation}
{\cal A}_\mu^c(x) = \frac{2\rho^2 \,\bar\eta^c_{\mu\nu}(x -
x_0)^\nu}{ (x - x_0)^2 \Big[(x-x_0)^2 + \rho^2\Big]}\simeq
\frac{2\rho^2 \,\bar\eta^c_{\mu\nu}\,(x - x_0)^\nu}{ (x -
x_0)^4}\left(1 - {\frac{\rho^2}{(x-x_0)^2}} + \ldots\right)~.
\label{solution1}
\end{equation}
This result explicitly shows that mixed disk diagrams, like that of Fig.
\ref{fig:md2}, are the source for the classical gauge instanton.
Note that the amplitude (\ref{dia1}) is a 3-point function from
the point of view of the two dimensional conformal field theory on
the string world sheet, but is a 1-point function from the point
of view of the four-dimensional gauge theory on the D3 branes. In
fact, the two boundary changing operators $V_{\bar w}$ and $V_{w}$
that appear in (\ref{dia1}) just describe non-dynamical parameters
on which the background depends. Furthermore, the fact that the
gluon field (\ref{solution}) is in the singular gauge is not
surprising, because in our set-up the gauge instanton is produced
by a D$(-1)$ brane which is a point-like object inside the D3
brane world-volume. Thus it is natural that the gauge connection
exhibits a singularity at the location $x_0$ of the D-instanton.

An obvious question at this point is whether also the subleading
terms in the large distance expansion (\ref{solution1}) have a
direct interpretation in string theory. Since such terms contain
higher powers of $\rho^2\sim \bar w^{\dot\alpha} w_{\dot\alpha}$,
one expects that they are associated to mixed disks with more
insertions of boundary changing operators. This expectation has
been explicitly confirmed in Ref.~\cite{Billo:2002hm} , so that one
can conclude that  mixed disks with the emission of a gauge vector
field do indeed reproduce the complete $k=1$ instanton
solution.

\section{Deformed $\mathcal{N}=2$ instanton calculus}
\label{sec:n2}

In this section we analyze the instanton moduli space of $\mathcal{N}=2$ gauge theories in
a non-trivial supergravity background. In particular we turn on a (self-dual) field
strength for the graviphoton of the $\mathcal{N}=2$ supergravity multiplet and see how it
modifies the instanton moduli action. This graviphoton background breaks Lorentz
invariance in space-time (leaving the metric flat) but it allows to explicitly perform
instanton calculations and establish a direct correspondence with the localization
techniques that have been recently discussed in the
literature~ \cite{Nekrasov:2002qd,Flume:2002az,Losev:2003py} .

In order to systematically incorporate the gravitational
background in the instanton action, let us first discuss how to
include the interactions among the instanton moduli and the gauge
fields. Then, let us consider all correlators involving {D3/D3} fields, and in particular
the scalar ${\phi}$ in presence of {$k$ D-instantons}. It turns
out~ \cite{Green:1997tv,Green:2000ke,Billo:2002hm}
that the dominant contribution to
the $n$-point function $\langle{\phi_1}\ldots {\phi_n}\rangle$ is from
{$n$ one-point} amplitudes on disks with moduli insertions.
The result can therefore be encoded in extra moduli-dependent vertices for {$\phi$'s},
{\it i.e.} in
{extra terms} in the moduli action containing such {one-point} functions
\begin{equation}
 \mathcal{S}_{\rm mod}({\phi};{\mathcal{M}_{(k)}}) =
{\phi}({x}) J_\phi({\widehat{\mathcal{M}}_{(k)}}) +
\mathcal{S}_{\rm mod}({\widehat{\mathcal{M}}_{(k)}})~,
\label{smod1}
\end{equation}
where {$x$} is the instanton center (previously denoted by $x_0$) and
${\phi}({x}) J_\phi({\widehat{\mathcal{M}}_{(k)}})$
is given by the disk diagrams with boundary (partly) on the D(--1)'s describing
the emission of a $\phi$.
To determine the complete action
$\mathcal{S}_{\mbox{\tiny mod}}({\phi};{\mathcal{M}})$
we have to systematically compute all mixed disks with a scalar
{$\phi$} emitted from the D3 boundary. Other non-zero diagrams involving
the instanton moduli and the super-partners of $\phi$ can be
obtained using the Ward identities of the supersymmetries that are broken by the D(--1) branes.
Therefore, the complete superfield-dependent moduli action
$\mathcal{S}_{\rm mod}({\Phi};{\mathcal{M}_{(k)}})$
can be obtained from (\ref{smod1}) by simply letting
${\phi(x)} \rightarrow {\Phi}({x},\theta)$, with $\Phi(x,\theta)$
defined in \eq{Pm1}.

Let us now extend this argument to the supergravity background we
want to include.
The field content of {$\mathcal{N}=2$ supergravity}, namely the metric
$h_{\mu\nu}$, the gravitini $\psi_\mu^{\alpha A}$ and the graviphoton $C_\mu$
can be organized in a {chiral Weyl multiplet}
\begin{equation}
{W^+_{\mu\nu}(x,\theta)}= {\mathcal{F}_{\mu\nu}^+(x)} +
\theta {\chi_{\mu\nu}^+(x)}+\frac{1}{2}
\,\theta\sigma^{\lambda\rho} \theta\,{R^+_{\mu\nu\lambda\rho}(x)}
+\cdots
\label{weyl}
\end{equation}
where the self-dual tensor ${\mathcal{F}_{\mu\nu}^+(x)}$
can be identified on-shell with the graviphoton field
strength, $R^+_{\mu\nu\lambda\rho}(x)$ is the self-dual Riemann curvature tensor and
$\chi_{\mu\nu}$ is the gravitino field strength.
All these fields belong to the massless sector of {type IIB strings} on
${\mathbb{R}^{4}}\times \mathbb{C}\times {\mathbb{C}^2/\mathbb{Z}_2}$.
In particular, the graviphoton vertex
is given by%
\footnote{A {different} R-R field, with a {similar} structure, will be
useful:
\begin{equation*}
{V_{\bar{\mathcal{F}}}}(z,\bar z) \! = \!
\frac{1}{4\pi}
{\bar{\mathcal{F}}^{\alpha\beta \hat A\hat B}}(p)
\Big[S_\alpha(z)S^+(z)S^{\dot A}(z)\ee^{-\frac{1}{2}\varphi(z)}
{S}_\beta(\bar z)S^+(z){S}^{\dot B}(\bar z)\ee^{-\frac{1}{2}{\varphi}(\bar
z)}\Big]\ee^{\ii p\cdot X(z,\bar z)}~.
\label{Vbarf}
\end{equation*}
}
\begin{equation}
{V_{\mathcal{F}}}(z,\bar z)  \!=\!  \frac{1}{4\pi}{\mathcal{F}^{\alpha\beta AB}}(p)
\Big[S_\alpha(z)S^-(z)S_A(z)\ee^{-\frac{1}{2}\varphi(z)}
{S}_\beta(\bar z)S^-(z){S}_B(\bar z)\ee^{-\frac{1}{2}{\varphi}(\bar
z)}\Big]\ee^{\ii p\cdot X(z,\bar z)}
\end{equation}
where the left-right movers identification on disks has already been taken
into account.
The bi-spinor graviphoton polarization is given by
\begin{equation}
{\mathcal{F}^{(\alpha\beta) [AB]}} = \frac{\sqrt 2}{4}\,
{\mathcal{F}_{\mu\nu}^+}\big(\sigma^{\mu\nu})^{\alpha\beta}\,\epsilon^{AB}
\end{equation}
and corresponds to a R-R 3-form $\mathcal{F}_{\mu\nu z}$ with one index in the
$\mathbb{C}$ internal direction.
To determine the contribution of the graviphoton to the field-dependent
moduli action we have to consider disk amplitudes with open string
moduli vertices on the boundary and closed string graviphoton
vertices in the interior, which survive in the field theory limit
$\alpha'\to 0$.

\begin{figure}[t]
\begin{center}
\psfrag{f}{\small $\bar{\mathcal{F}}^+$}
\psfrag{mm}{\small $M$}
\psfrag{m}{\small $M$}
\includegraphics[width=0.18\textwidth]{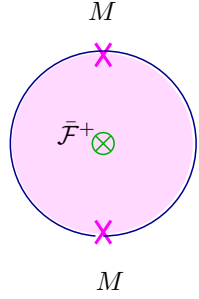}
\end{center}
\caption{Mixed disk describing the coupling among the instanton moduli and a closed
string field.}\label{fig:3}
\end{figure}

It turns out that very few diagrams (like the one represented in Fig. \ref{fig:3})
contribute in this limit. They can be easily evaluated and for instance for the diagram in
Fig. \ref{fig:3} one finds
\begin{equation}
\lvev {V_{M} V_{M}}{V_{\bar{\mathcal F}}} \rvev
=\frac{1}{4\sqrt 2}\mbox{tr}_k\Big\{{M^{\alpha A}M^{\beta B}}
{\bar{\mathcal{F}}^+_{\mu\nu}}
\Big\}(\sigma^{\mu\nu})_{\alpha\beta}\epsilon_{AB}~.
\label{mmbare}
\end{equation}
Other diagrams, connected by the broken supersymmetries,
have the effect of promoting the dependence of the moduli action to the full Weyl
multiplet, {\it i.e.}
${\mathcal{F}^+_{\mu\nu}} \to {W^+_{\mu\nu}}(x,\theta)$.
In this way the superfield-dependent moduli action
$\mathcal{S}_{\rm mod}({\Phi},{W^+};{\widehat{\mathcal{M}}_{(k)}})$
is obtained from the previous results.

Integrating over the moduli,
one gets the effective action, and hence the prepotential at instanton number
$k$, namely
\begin{equation}
\label{effacW}
\begin{aligned}
 S_{\rm eff}^{(k)}[{\Phi},{W^+}] &=\int d^4x \, d^4\theta\,\,
d{\widehat{\mathcal{M}}_{(k)}}\, \ee^{-\frac{8\pi k}{g^2} -
\mathcal{S}_{\rm mod}({\Phi},{W^+};{\widehat{\mathcal{M}}_{(k)}})}
\\
&=\int d^4x \, d^4\theta\,\,\, \mathrm{F}^{(k)}({\Phi},{W^+})
~.
\end{aligned}
\end{equation}
Since {$\Phi(x,\theta)$} and {$W_{\mu\nu}^+(x,\theta)$} are constant
with respect to the integration variables ${\widehat{\mathcal{M}}_{(k)}}$,
we can actually compute $\mathrm{F}^{(k)}$ by reducing them to a
constant value, {\it i.e.} ${\Phi}(x,\theta) \to {a}$ and
${W^+_{\mu\nu}}(x,\theta) \to {f_{\mu\nu}}$.
In this case the prepotential becomes just a function of the scalar and graviphoton
v.e.v.'s and is determined by
a ``deformed'' moduli action depending on $a, {\bar a}, f, {\bar f}$:
\begin{eqnarray}
\label{defmodac}
&&\hskip -0.2cm
\mathcal{S_{\rm mod}}({a,\bar a};{f,\bar f};{\widehat{\mathcal{M}}_{(k)}}) =
- \mbox{tr}_k\Big\{
\big([\chi^{\dagger},a'_{\alpha\dot\beta}]+2{\bar f_c}
(\tau^c a')_{\alpha\dot\beta}\big)
\big([\chi,{a'}^{\dot\beta\alpha}]+2{f_{c}}(a'\tau^c)^{\dot\beta \alpha}\big)
\nonumber \\
&&\hskip -0.2cm
-\big(\chi^{\dagger}{\bar w}_{\dot\alpha}-{\bar w}_{\dot\alpha}\,{\bar a}\big)
\big( w^{\dot\alpha}\chi- {a}\,w^{\dot\alpha}\big)
-\big(\chi{\bar w}_{\dot\alpha} -{\bar w}_{\dot\alpha}\,{a}\big)
\big(w^{\dot\alpha}\chi^{\dagger}
-{\bar a}\,w^{\dot\alpha}\big) \Big\}
\\
&&\hskip -0.2cm
+\ii\, \frac{\sqrt 2}{2}\,
\mbox{tr}_k\Big\{{\bar \mu}^A \epsilon_{AB} \big( \mu^B\chi^{\dagger}
-{\bar a}\,\mu^B\big)
-\frac{1}{2}\,M^{\alpha A}\epsilon_{AB}\big([\chi^{\dagger},M_{\alpha}^{B}]
+2\,{\bar f_c}\, (\tau^c)_{\alpha\beta}M^{\beta B}\big)\Big\}
\nonumber \\
&&\mbox{tr}_k\Big\{-\ii  D_c \big({w_{\dot\alpha}(\tau^c)^{\dot\alpha}_{\dot\beta}
\bar{w}_{\dot\beta}
 +\ii
\bar\eta_{\mu\nu}^c \big[{a'}^\mu,{a'}^\nu\big]}\big) 
+ \ii {\lambda}^{\dot\alpha}_{\,A}\big(
{\bar{\mu}^A{w}_{\dot\alpha}+
\bar{w}_{\dot\alpha}{\mu}^A  +
\big[a'_{\alpha\dot\alpha},{M'}^{\alpha A}\big]\big)}\Big\}~.
\nonumber
\end{eqnarray}
Notice that the ADHM constraints, appearing in the last line
of (\ref{defmodac}), are not modified by the graviphoton background.

In the action (\ref{defmodac})
the v.e.v.'s ${a},{f}$ and ${\bar a},{\bar f}$ are not on the same footing.
Indeed, one can write
\begin{equation}
\mathcal{S}_{\rm mod}({a,\bar a};{f,\bar f};{\widehat{\mathcal{M}}_{(k)}}) = {Q}\,\Xi
\end{equation}
where {$Q$} is the scalar component
of the twisted supercharges, {\it i.e.}
\begin{equation}
\label{Qscalar}
{Q} \equiv
\frac{1}{2}\,\epsilon_{\dot\alpha\dot\beta}\,Q^{\dot\alpha\dot\beta}~,
\end{equation}
where the topological twist acts as
$Q^{\dot\alpha B}\stackrel{\mbox{\tiny top. twist}}{\longrightarrow}
Q^{\dot\alpha\dot\beta}$. It turns out that the parameters $\bar a$, $\bar f$ appear in
$\mathcal{S}_{\rm mod}$ only through the gauge fermion $\Xi$,
and thus the instanton partition function
and the prepotential $\mathrm{F}^{(k)}$ in \eq{effacW} are in fact
independent of $\bar a$,$\bar f$, because their variation with
respect to these parameters is {$Q$-exact}.

{F}rom the explicit expression of $\mathcal{S_{\rm mod}}({a,0};{f,0})$
the general form of the prepotential $\mathrm{F}^{(k)}({a};{f})$
can be easily deduced; reinstating the superfields it reads
\begin{equation}
\label{Fkpw}
\mathrm{F}^{(k)}({\Phi}, {W^+}) =  \sum_{h=0}^\infty {c_{k,h}} \,
{\Phi^2}
\left(\frac{\Lambda}{{\Phi}}\right)^{4k}\!\left(\frac{{W^+}}
{{\Phi}}\right)^{2h}~.
\end{equation}
Summing over the instanton sectors we obtain the full non-perturbative
prepotential
\begin{equation}
\label{Fnp}
\mathrm{F}_{\mbox{\tiny n.p.}}({\Phi},{W^+}) =
\sum_{k=1}^\infty \mathrm{F}^{(k)}({\Phi},{W^+})
= \sum_{{h}=0}^\infty C_{{h}}(\Lambda,{\Phi}) {(W^+)^{2h}}~,
\end{equation}
where
\begin{equation}
\label{Chis}
C_{{h}}(\Lambda,{\Phi})
=
\sum_{k=1}^\infty {c_{k,h}} \,\frac{\Lambda^{4k}}{{\Phi}^{4k+2{h}-2}}~.
\end{equation}
This gives rise to many different terms in the effective action,
which is obtained, see \eq{effacW},
by integrating the prepotential over $d^4x\, d^4\theta$. In particular,
saturating the $\theta$ integration with
four $\theta$'s all from ${W^+}$ we get gravitational
F-terms in the $\mathcal{N}=2$ effective action involving the curvature tensor and
graviphoton field strength
\begin{equation}
\label{R2W}
\int d^4x
\,\,C_{{h}}(\Lambda,{\phi})\,({R^+})^2  ({\mathcal{F}^{+}})^{2h-2}~.
\end{equation}
The stringy instanton calculus accounts thus for such F-terms, and it gives
a way to compute them, because the coefficients $C_{k,h}$ can be explicitly
determined by performing the integrals over the instanton moduli space.

This is a formidable task, that was finally performed in Refs.~
\cite{Nekrasov:2002qd,Flume:2002az,Losev:2003py} ,
using a suitable ``deformation'' of the moduli action
which localizes the integrals. This localization deformation exactly
coincides with \eq{defmodac} if we set
\begin{equation}
\label{floc}
{f_c}=\frac{{\varepsilon}}{2}\,\delta_{3c}~,~~
{\bar f_c} = \frac{{\bar\varepsilon}}{2}\,\delta_{3c}~,~~
\end{equation}
(and moreover ${\varepsilon}={\bar \varepsilon}$).

As we remarked above, $Z^{(k)}({a},{\varepsilon})$ does not depend on
{$\bar\varepsilon$}.
However, ${\bar\varepsilon}=0$ is a limiting case and some care is needed.
In fact, while $\mathrm{F}^{(k)}({a};{\varepsilon})$ is
well-defined, the complete partition function
$Z^{(k)}({a};{\varepsilon})$ diverges because of the (super)volume integral
$\int d^4x \,d^4\theta$. The presence of
{$\bar\varepsilon$} regularizes the superspace integration by a Gaussian term,
leading to the following effective rule:
\begin{equation}
\label{epsirule}
\int d^4x\, d^4\theta \to 1/\varepsilon^2~;
\end{equation}
one can then work with the \emph{full} instanton partition function.
Moreover, the $a$ and $\varepsilon,\bar\varepsilon$ deformations localize
completely the integration over moduli space which can then be
evaluated explicitly~ \cite{Nekrasov:2002qd,Flume:2002az,Losev:2003py} .

With $\bar\varepsilon\not=0$, {\it i.e.} with complete localization, the trivial
superposition of several instantons of charges $k_i$ contributes to the sector $k= \sum k_i$;
such disconnected configurations do \emph{not} contribute instead
when $\bar\varepsilon=0$.
The partition function computed by localization thus corresponds in this case
to the exponential of the non-perturbative prepotential, namely
\begin{equation}
\label{ZvsF}
\begin{aligned}
Z({a};{\varepsilon}) &= \sum_{k=1}^\infty Z^{(k)}({a};{\varepsilon})  =
\exp\left(\frac{\mathcal{F}_{\mbox{\tiny n.p.}}({a},
{\varepsilon})}{{\varepsilon^2}}\right)
= \exp\left(\sum_{k=1}^\infty \frac{\mathcal{F}^{(k)}({a},
{\varepsilon})}{{\varepsilon^2}}\right)
\\
& =  \exp\left(\sum_{h=0}^\infty \sum_{k=1}^\infty {c_{k,h}}
\frac{{\varepsilon^{2h-2}}}{{a^{2h-2}}}
\left(\frac{\Lambda}{{a}}\right)^{4k}
\right)~.
\end{aligned}
\end{equation}
In conclusion, the computation via localization techniques of the
multi-instanton partition function
$Z({a};{\varepsilon})$ determines the coefficients ${c_{k,h}}$
which appear in the gravitational F-terms of the $\mathcal{N}=2$
effective action \eq{R2W}
via the expression of $C_{{h}}(\Lambda,{\phi})$ given in \eq{Fnp}.

The very same gravitational F-terms can been extracted in a completely
different way by considering topological string amplitudes at genus $h$
on suitable Calabi-Yau manifolds~ \cite{Bershadsky:1993cx,Antoniadis:1993ze} .
In our computation the role of the genus $h$ Riemann surface is played
by a (degenerate) surface with the same Euler number made by $2h$
disconnected disks, instead of $h$ handles.
The two different roads to determine the $F$-couplings of \eq{R2W} must
lead to the same result.
This is a very natural way to state the conjecture by N. Nekrasov~
\cite{Nekrasov:2002qd} that the coefficients arising in the
$\varepsilon$-expansion of multi-instanton partition functions match
those appearing in higher genus topological string amplitudes on Calabi-Yau manifolds.

\section*{Acknowledgements}
We would like to thank F. Fucito and I. Pesando for many useful
discussions. We thank the Galileo Galilei Institute for
hospitality and support. This work is partially supported by the European Commission FP6
Programme MRTN-CT-2004-005104, in which A.L. is associated to University of Torino, and by
Italian MUR under contract PRIN-2005023102.


\begin{thebibliography}{99}

\bibitem{Witten:1995im} E.~Witten,
Nucl.\ Phys.\ {\bf B460} (1996) 541
[arXiv:hep-th/9511030].

\bibitem{douglas}
M.~R.~Douglas,
``Branes within branes'' [arXiv:hep-th/9512077];
J.\ Geom.\ Phys.\ {\bf 28}, 255 (1998)
[arXiv:hep-th/9604198];

\bibitem{Green:2000ke}
M.~B.~Green and M.~Gutperle,
JHEP {\bf 0002} (2000) 014
[arXiv:hep-th/0002011].

\bibitem{Dorey:2002ik}
N.~Dorey, T.~J.~Hollowood, V.~V.~Khoze and M.~P.~Mattis,
Phys.\ Rept.\  {\bf 371} (2002) 231
[arXiv:hep-th/0206063].
M.~Bianchi, S.~Kovacs and G.~Rossi,
``Instantons and supersymmetry,''[arXiv:hep-th/0703142].

\bibitem{Billo:2002hm}
M.~Billo, M.~Frau, I.~Pesando, F.~Fucito, A.~Lerda and A.~Liccardo,
JHEP {\bf 0302} (2003) 045
[arXiv:hep-th/0211250];
M.~Frau and A.~Lerda,
Fortsch.\ Phys.\ {\bf 52} (2004) 606
[arXiv:hep-th/0401062].

\bibitem{Billo:2005fg}
M.~Billo, M.~Frau, S.~Sciuto, G.~Vallone and A.~Lerda,
JHEP {\bf 0605} (2006) 069
[arXiv:hep-th/0511036].

\bibitem{Billo:2004zq}
M.~Billo, M.~Frau, I.~Pesando and A.~Lerda,
JHEP {\bf 0405} (2004) 023
[arXiv:hep-th/0402160];
M.~Billo, M.~Frau, F.~Lonegro and A.~Lerda,
JHEP {\bf 0505} (2005) 047
[arXiv:hep-th/0502084].

\bibitem{recent} R.~Blumenhagen, M.~Cvetic and T.~Weigand,
  Nucl.\ Phys.\  B {\bf 771} (2007) 113
  [arXiv:hep-th/0609191].
  L.~E.~Ibanez and A.~M.~Uranga,
  JHEP {\bf 0703} (2007) 052
  [arXiv:hep-th/0609213].
  B.~Florea, S.~Kachru, J.~McGreevy and
  N.~Saulina, 
  arXiv:hep-th/0610003.  
  S.~A.~Abel and M.~D.~Goodsell,
  arXiv:hep-th/0612110.
  N.~Akerblom, R.~Blumenhagen, D.~Lust, E.~Plauschinn and M.~Schmidt-Sommerfeld,
  JHEP {\bf 0704} (2007) 076
  [arXiv:hep-th/0612132].
  M.~Bianchi and E.~Kiritsis,
  arXiv:hep-th/0702015.  
  M.~Cvetic, R.~Richter and T.~Weigand,
  arXiv:hep-th/0703028.  
  R.~Argurio, M.~Bertolini, G.~Ferretti, A.~Lerda and C.~Petersson,
  arXiv:0704.0262 [hep-th].
  M.~Bianchi, F.~Fucito and J.~F.~Morales,
  arXiv:0704.0784 [hep-th].
  L.~E.~Ibanez, A.~N.~Schellekens and A.~M.~Uranga,
  arXiv:0704.1079 [hep-th].

\bibitem{Billo:2006jm}
M.~Billo, M.~Frau, F.~Fucito and A.~Lerda,
JHEP {\bf 0611} (2006) 012
[arXiv:hep-th/0606013].

\bibitem{Nekrasov:2002qd}
N.~A.~Nekrasov,
Adv.\ Theor.\ Math.\ Phys.\  {\bf 7} (2004) 831
[arXiv:hep-th/0206161].

\bibitem{Losev:2003py}
A.~S.~Losev, A.~Marshakov and N.~A.~Nekrasov,
``Small instantons, little strings and free fermions''
[arXiv:hep-th/0302191];
N.~Nekrasov and A.~Okounkov,
``Seiberg-Witten theory and random partitions''
[arXiv:hep-th/0306238].

\bibitem{Fucito:1996ua}
F.~Fucito and G.~Travaglini,
Phys.\ Rev.\ D {\bf 55} (1997) 1099
[arXiv:hep-th/9605215].

\bibitem{Hollowood:2002ds}
T.~J.~Hollowood,
JHEP {\bf 0203} (2002) 038
[arXiv:hep-th/0201075].

\bibitem{Antoniadis:1993ze}
I.~Antoniadis, E.~Gava, K.~S.~Narain and T.~R.~Taylor,
Nucl.\ Phys.\ B {\bf 413} (1994) 162
[arXiv:hep-th/9307158].

\bibitem{Flume:2002az}
R.~Flume and R.~Poghossian,
Int.\ J.\ Mod.\ Phys.\ A {\bf 18} (2003) 2541
[arXiv:hep-th/0208176].
R.~Flume, F.~Fucito, J.~F.~Morales and R.~Poghossian,
JHEP {\bf 0404} (2004) 008
[arXiv:hep-th/0403057].

\bibitem{Green:1997tv}
M.~Green and M.~Gutperle,
Nucl.\ Phys.\ {\bf B498} (1997) 195
[arXiv:hep-th/9701093];
JHEP {\bf 9801} (1998) 005
[arXiv:hep-th/9711107];
Phys.  Rev.  {\bf D58} (1998) 046007
[arXiv:hep-th/9804123].

\bibitem{Bershadsky:1993cx}
  M.~Bershadsky, S.~Cecotti, H.~Ooguri and C.~Vafa,
  Commun.\ Math.\ Phys.\  {\bf 165} (1994) 311
  [arXiv:hep-th/9309140].
\end{thebibliography}
\end{document}